\documentclass[iop,apj]{emulateapj}

\usepackage{amsmath,amssymb,amstext}

\usepackage[breaklinks,colorlinks,citecolor=blue,linkcolor=magenta]{hyperref} 

\usepackage[all]{hypcap} 

\usepackage{aas_macros}
\usepackage{natbib}

\usepackage{booktabs}
\usepackage{array}
\usepackage{arydshln}
\usepackage{todonotes}
\newcolumntype{L}{>{\centering\arraybackslash}m{1.5cm}}  
\newcolumntype{M}{>{\centering\arraybackslash}m{2.5cm}}

\shorttitle{Rocky Planets Ejected During Late-stage Accretion}
\shortauthors{Barclay et al.} 

\begin{document}

\title{The Demographics of Rocky Free-Floating Planets and their Detectability by WFIRST}

\author{Thomas Barclay\altaffilmark{1,2}, Elisa V. Quintana\altaffilmark{1}, Sean N. Raymond\altaffilmark{3}, and Matthew T. Penny\altaffilmark{4,5}}
\email{thomas.barclay@nasa.gov}
\altaffiltext{1}{NASA Goddard Space Flight Center, 8800 Greenbelt Road, Greenbelt, MD 20771, USA}
\altaffiltext{2}{University of Maryland, Baltimore County,1000 Hilltop Cir, Baltimore, MD 21250, USA}
\altaffiltext{3}{Laboratoire d'astrophysique de Bordeaux, Univ. Bordeaux, CNRS, B18N, all\'{e}e Geoffroy Saint-Hilaire, F-33615 Pessac, France}
\altaffiltext{4}{Department of Astronomy, Ohio State University, 140 West 18th Avenue, Columbus, OH 43210, USA}
\altaffiltext{5}{Sagan Fellow}

\begin{abstract}
Planets are thought to form via accretion from a remnant disk of gas and solids around a newly formed star. During this process material in the disk either remains bound to the star as part of either a planet, a smaller celestial body, or makes up part of the interplanetary medium; falls into the star; or is ejected from the system. Herein we use dynamical models to probe the abundance and properties of ejected material during late-stage planet formation and estimate their contribution to the free-floating planet population. We present 300 $N$-body simulations of terrestrial planet formation around a solar-type star, with and without giant planets present, using a model that accounts for collisional fragmentation. In simulations with Jupiter and Saturn analogs, about one-third of the initial ($\sim$5 $M_\oplus$) disk mass is ejected, about half in planets more massive than Mercury but less than 0.3 $M_\oplus$, and the remainder in smaller bodies. Most ejections occur within 25 Myr, which is shorter than the timescale typically required for Earth-mass planets to grow (30--100 Myr). When giant planets are omitted from our simulations, almost no material is ejected within 200 Myr and only about 1\% of the initial disk is ejected by 2 Gyr. We show that about 2.5 terrestrial-mass planets are ejected per star in the Galaxy. We predict that the space-borne microlensing search for free-floating planets from the Wide-Field Infra-Red Space Telescope (WFIRST) will discover up to 15 Mars-mass planets, but few free-floating Earth-mass planets. 

\end{abstract}


\maketitle

\section{Introduction}
In the classical picture of planet formation, planets grow via accretion from material within a protoplanetary disk that remains around a newly formed star. During this process, material is either accreted by planets or smaller bodies, resides in the interplanetary medium, falls into the central star, or is imparted with enough angular momentum to be ejected from the system. So-called free-floating planets (or rogue planets, wandering planets, etc.) have been observed by gravitational microlensing surveys \citep{bennett13,freeman15} and by optical and infrared wide-field surveys \citep{lucas00,bihain09,delorme12,lui13,dupuy13}. Estimates have been made that there are as many as 2 free-floating Jupiter-mass planets for every star in the Galaxy \citep{sumi11,clanton16}. Although it is unlikely that all the giant free-floating planets were ejected from planetary systems during planet formation \citep{veras12,chatterjee08,pfyffer15,ma2016} because this requires a higher occurrence rate of bound giant planets than is observed, it is probable that ejected planets make up some fraction of the free-floating population. It is also possible that binary stars produce the majority of the free-floating giant planet population \citep{sutherland2016,smullen2016}.

Hitherto, the detected free-floating planets have been giant worlds that could potentially represent the tail-end of the stellar mass distribution \citep{silk77}. However, the microlensing experiment \citep{henderson15} from the K2 mission \citep{howell14} is potentially sensitive to super-Earth-mass bodies and above \citep{henderson16b,penny2016}. The Wide-Field Infra-Red Survey Telescope (WFIRST) from NASA, due to launch in 2024 \citep{spergel15}, will be sensitive to bodies less massive than Earth, and the ESA mission Euclid will be sensitive to sub-Earth-mass planets \citep{penny13}.

Ejected material is a natural outcome of the planet formation process. The process begins when a molecular cloud collapses to form a star that is subsequently surrounded by a remnant disk of gas and dust. The growth of planets from within this disk occurs in several distinct phases that are typically addressed independently due to their different physical processes. In the early phase, dusty material coalesces into 1-10 km-sized planetesimals in a process that isn't yet fully understood. The planetesimals accrete material within their gravitational zone of influence and form embryos that are typically Mars-mass in the terrestrial region \citep{Lissauer.Stewart:1993,Kokubo.Ida:2000,leinhardt05}. Up to this point, gas is still present in the disk and massive cores can accrete gaseous envelopes to form giant planets \citep{Lissauer:1987,Lissauer:2009,papaloizou06}. In the final stages of planet formation, the gas in the disk is dispersed and the process becomes dominated by gravitational interactions and collisions. The solid material in the disk is stirred as bodies grow or scatter off each other, and this process continues for tens of millions of years until final planets form on stable, widely separated orbits -- the so-called `giant impact phase' \citep{Morbidelli.etal:2012,Raymond.Morbidelli:2014}. This description applies to Sun-like stars, but it is expected that similar processes occur for planets forming around other star types, albeit on different timescales.

If gas giants are present while the terrestrial planets are forming, as was the case in our Solar System, their presence will dominate the dynamics and excite the eccentricities and inclinations of the protoplanets in the disk \citep{levison03,chambers02b,Raymond.etal:2006b}. The degree of excitation is sensitive to the architecture of the giant planets. Nearby, more massive and/or more eccentric giant planets will cause stronger perturbations that can lead to crossing orbits among the bodies and ultimately to ejections.

Numerical $N$-body models have been widely used to study terrestrial planet formation from a range of initial conditions -- star types, disk mass distributions, etc. \citep{Wetherill:1994,Chambers:2001,Raymond.etal:2004,Raymond.etal:2006a,Raymond.etal:2006b,Raymond.etal:2009,Quintana.etal:2002,Quintana.Lissauer:2006,Quintana.etal:2010,Quintana.Lissauer:2014,morishima08,morishima10,Fischer.Ciesla:2014,ciesla15,kominami04,ogihara07}. While a large number of studies have been performed, details of the ejected material from these studies is typically not well documented, as the focus of these studies has been to form and characterize terrestrial planets and not necessarily to track the fate of ejected mass. In some cases the fraction of disk mass that was ejected during a simulation is reported \citep{Quintana.etal:2002,Quintana.Lissauer:2006,Quintana.etal:2010,Quintana.Lissauer:2014}, but the size distribution and timescales of the ejected material has not been analyzed. Regardless, the number of realizations per system has typically not been large enough to draw statistically meaningful conclusions.
 
Historically, these $N$-body accretion models have been limited in two key ways, both attributed to the computationally intensive nature of these types of simulations. First, collisions have been treated as perfect mergers, meaning two bodies that collide stick together and conserve mass and momentum. Second, a relatively small number of realizations have typically been performed for a given star/disk configuration. The recent development of an analytical prescription for collision outcomes in the gravity-dominated regime \citep{Leinhardt.Stewart:2012,Stewart.Leinhardt:2012} provided a feasible way to realistically model `hit-and-run' events \citep{Asphaug.etal:2006} and fragmentation from energetic impacts \citep[e.g.][]{,Stewart.etal:2015}. The model was incorporated into the popular \emph{Mercury} $N$-body integration package \citep{Chambers:1999,Chambers:2013}. \citet{quintana16} showed using this version of \emph{Mercury} that the inclusion of fragmentation produced a dramatic improvement in studying the aspects of planet formation that were sensitive to the fate and evolution of material compared with older work that assumed perfect accretion. The second recent innovation was to utilize the NASA Pleiades Supercomputer to address the problem of small sampling of highly stochastic processes (such as these $N$-body accretion models) by performing hundreds of simultaneous simulations with near-identical initial conditions \citep{quintana16}. These improvements allow for probabilistic predictions based on a relatively large number of samples as well as quantifying the occurrence rates of somewhat uncommon outcomes.

In this work we present the results from 300 $N$-body simulations of late-stage terrestrial planet formation around a Sun-like star that we performed using our fragmentation model. Half of these simulations include giant planets analogous to Jupiter and Saturn, while half lack giant planet companions. Our simulations examine growth from a disk of hundreds of protoplanets within 4 AU from the star, and we track the fate of all bodies as the systems are evolved for 2 Gyr. We quantify the ejected material to make a prediction on the mass distribution of free-floating planets that result from the planet formation process. 

We also analyze results from a suite of 152 simulations performed by \citet{raymond11, raymond12} that began with a disk of protoplanets that extended out to 10 AU and included three giant planets to estimate the abundance of ejected terrestrial-mass planet in systems with unstable giant planets.  

Finally, we discuss the implications these results have for the WFIRST microlensing experiment.


\section{Numerical Model}\label{sec:model}
The simulations described here were performed to explore the fate of planets that formed from material that originated within the terrestrial planet region (within 4 AU). We follow the accretion of solids during the final stages of the planet formation process at an epoch that corresponds to about 10 Myr after the start of the Solar System's formation \citep{quintana16}. At this epoch, tens of Mars-sized embryos are thought to have formed from solid material in the disk along with a large number of Moon-sized and smaller planetesimals. All gas in the disk has been dispersed, therefore giant planets, if included, are assumed to be fully formed. In 150 of our simulations we include giant planets analogous to Jupiter and Saturn on orbits comparable to their present locations. Another set of 150 simulations were performed without giant planets.

The growth of Mars-sized embryos is supported by simulations of earlier stages of planet formation \citep{Kokubo.Ida:1998} as well as the predicted isolation mass in theoretical planet formation models of our Solar System \citep{Lissauer.Stewart:1993}. We adopt the bimodal disk model from \citet{quintana16} in which half of the disk mass is comprised of 26 approximately Mars-mass bodies (0.0933 $M_\oplus$) and the other half is in 260 approximately lunar-mass bodies (0.00933 $M_\oplus$). 

The initial disk contains 4.85 $M_\oplus$ of solid material extending from 0.3 to 4 AU from a 1 $M_\odot$ central star. The initial bodies are spaced to fulfill a surface-density profile proportional to the semimajor axis raised to the $-3/2$ power, which results in bodies spaced by 3--6 Hill radii. This disk model is based on the `minimum mass Solar nebula' \citep{Weidenschilling:1977}, a model derived by essentially smoothing each of the eight planets in our Solar System into concentric rings and fitting a curve to estimate the surface-density profile of solid material that ultimately formed the planets in our Solar System. Disk models that follow this model with a bimodal mass distribution have been successful in numerically reproducing the broad characteristics of the terrestrial planets in our Solar System \citep{Chambers:2001,Raymond.etal:2004,Raymond.etal:2009,Chambers:2013,quintana16}. 

The small bodies are a proxy for what should ideally be millions of objects with masses ranging from about a lunar mass down to dust grains, where the mode of the distribution is thought to be at about 150 km sized bodies \citep{Bottke:2005,johnasen12,johansen15}. The resolution of 260 small bodies is chosen to keep the simulations computationally tractable (since for $N$-body models the computation time scales with the square of the number of bodies). Numerical simulations have shown that this resolution is sufficient to provide dynamical friction (i.e., the damping of eccentricities and inclinations of larger bodies due to the swarm of smaller bodies), which is an important mechanism to include in models of late-stage planet formation \citep{Chambers:2001,Obrien.etal:2006,Raymond:2006,morishima10}.

All simulations were evolved forward in time using the modified version of the $Mercury$ $N$-body integrator \citep{Chambers:1999} that accounts for collisional fragmentation \citep{Chambers:2013,quintana16}. If a planet is eroded during a two-body collision, the lost mass is broken up into \emph{fragments} that each have a mass equal to a chosen minimum fragmentation mass. The minimum mass for the fragments is necessary in order to constrain the total number of bodies in the integration. We use a value of 0.38 Moon-mass (0.005 $M_\oplus$) that was used in previous simulations \citep{Chambers:2013,quintana16} and shown to be computationally tractable. Our simulations assume a material density of 3 g/cm$^{3}$, which gives radii for the embryos, planetesimals and fragments of 0.56 $R_\oplus$ (3500 km), 0.26 $R_\oplus$ (1600 km), and 0.2 $R_\oplus$ (1300 km), respectively.

All simulations used virtually the same initial mass distribution for each disk, but each realization was minutely altered by perturbing a single body at approximately 1 AU by one meter in order to account for chaos. Of the 150 simulations with giant planets, 140 are the same simulations published by \citet{quintana16} with an additional 10 simulations performed for this work.

We allow these simulations to evolve for 2 Gyrs with a 7-day time-step allowing all initial bodies and fragments created during collisions to gravitationally interact. Any body that travels within 1 Solar radius of the host star is considered accreted by the central star. Any body that travels farther than 100 AU from the central star is deemed to be ejected from the system and is no longer tracked in the simulation. While a small fraction of these bodies that travel farther from their star than 100 AU may remain bound, this ejection distance was chosen because planets beyond this distance would be classified as free-floating planets to microlensing surveys such as the one WFIRST will perform \citep[e.g.][]{sumi11}. At each collision or ejection event, the masses, orbits, and collision parameters of the bodies involved are recorded. 

\section{Simulation Results}\label{sec:results}
A summary of the aggregate properties of material ejected is presented in Table~\ref{tab:results}. Following \citet{Quintana.Lissauer:2014}, we define a `planet' as one that grew at least as massive as the planet Mercury (0.06 $M_\oplus$). Although planets can be smaller than this \citep[e.g.][]{Barclay.etal:2013}, we keep this definition for consistency with previous work. A single initial large body or seven initial small bodies would satisfy the mass constraint for a planet. This definition allows us to distinguish between bodies whose number and mass are properly tracked in the simulation and contribute to our planet count, and the planetesimal and fragmented material that remain in the system. 

There are very dramatic differences between simulations with and without giant planets. With giant planets a total of 16320 bodies were ejected in the 150 simulations, of which 1192 were planets. This is an average of 7.9 large bodies per system. In contrast, not a single planet was ejected from systems without giant planets, and a total of just 1697 small bodies were ejected. 

\begin{figure}
\plotone{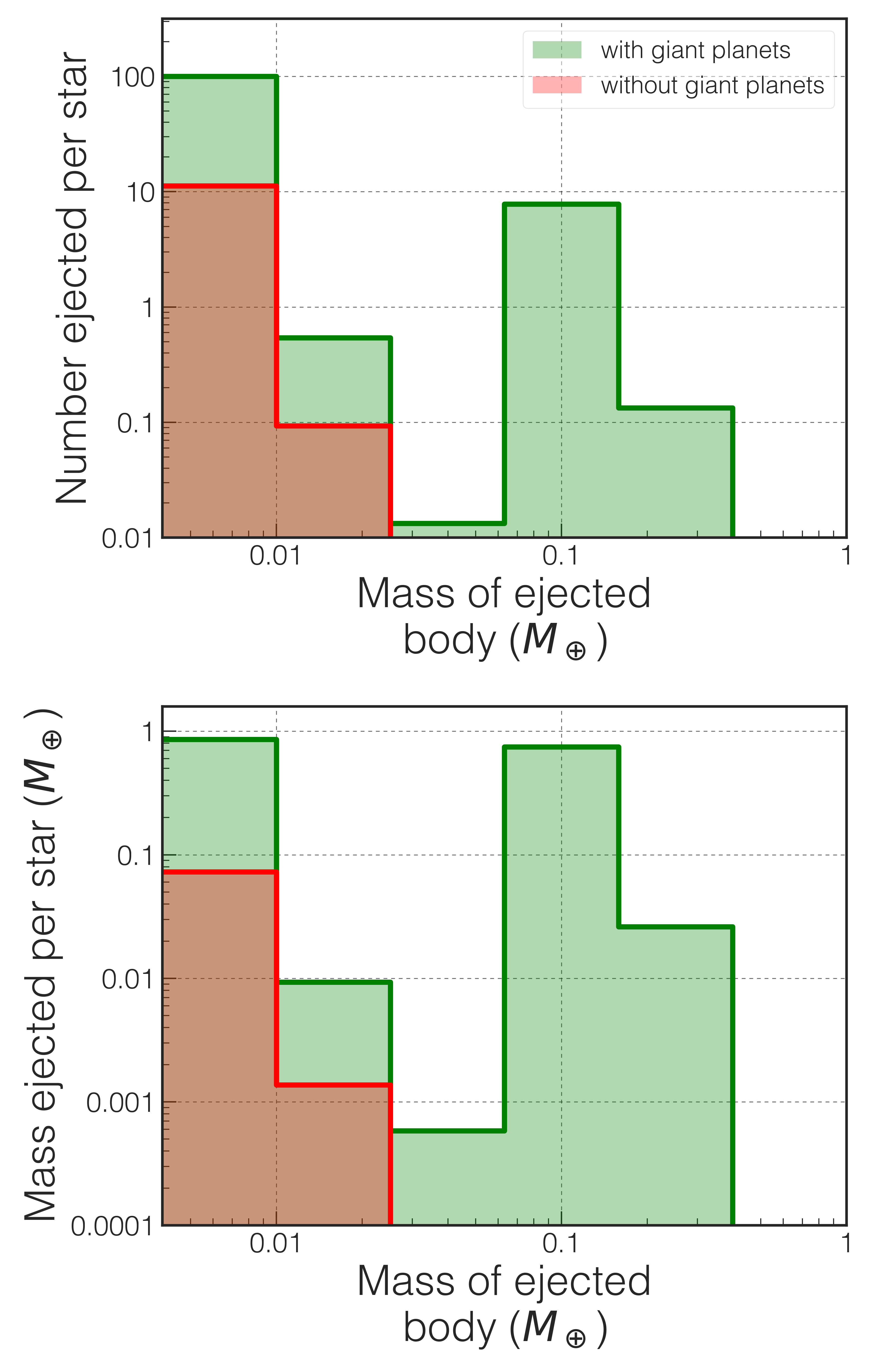}
\caption{The upper panel shows the number of bodies ejected per star in various mass bins for simulations with (green) and without (red) giant planets. The lower panel shows the ejected mass per star in each mass bin. With giant planets, about half the ejected mass is in planetesimals and half in planets, although many more small bodies are ejected. Without giant planets, far fewer planetesimals and no planets leave the system. Note that no bodies larger than 0.3 $M_\oplus$ were ejected from any of these simulations. 
\label{fig:ejectionmass}}
\end{figure}

With giant planets, an average of $1.64\pm0.15$ $M_\oplus$ of material of the initial $\sim$5 $M_\oplus$ disk was ejected during each 2 Gyr simulation. For simulations without giant planets, the central 90th percentile of the mass ejected ranges from 0.01--0.14 $M_\oplus$. In addition to the material ejected from the system, 1.0 $M_\oplus$ per simulation with giant planets, and 0.05 $M_\oplus$ per simulation without fell into the central star. The upper panel in Figure~\ref{fig:ejectionmass} shows the number of bodies ejected in various mass bins, normalized by the number of simulations. Far more small bodies are ejected than large bodies in simulations with giant planets (green shaded bars), but as shown in the lower panel of Figure~\ref{fig:ejectionmass} the ejected mass is distributed roughly equally among planets and planetesimals. Although many Mars-mass planets are ejected, no planets more than three Mars-masses are ever ejected from these simulations. Without giant planets (red shaded bars), both the number of ejected bodies and their masses are significantly lower compared to systems with giant planets.


\begin{table}
\caption{The properties of material ejected for our simulations}
\centering
\begin{tabular}{M L L} 
 \toprule
  & With giant planets& Without giant planets\\  
 \midrule
Average mass ejected per system& 1.64 $M_\oplus$& 0.07 $M_\oplus$\\
 \midrule
 Average mass ejected per system in planets& 0.77 $M_\oplus$& 0.00 $M_\oplus$\\
 \midrule
 Average mass ejected per system in planetesimals& 0.87 $M_\oplus$& 0.07 $M_\oplus$\\
 \midrule
 Average number of planets ejected & 7.9 & 0.0\\
  \midrule
 Average \% of initial disk ejected & 33 & 1 \\
 \midrule
 Average \% of ejected mass in planets & 47 & 0 \\
 \midrule
Average \% of ejected mass in planetesimals & 53 & 100\\ 
 \midrule
Average mass accreted by central star & 1.00 $M_\oplus$& 0.05 $M_\oplus$\\ 
 \bottomrule
\end{tabular}
\label{tab:results}
\end{table}

\begin{figure}
\plotone{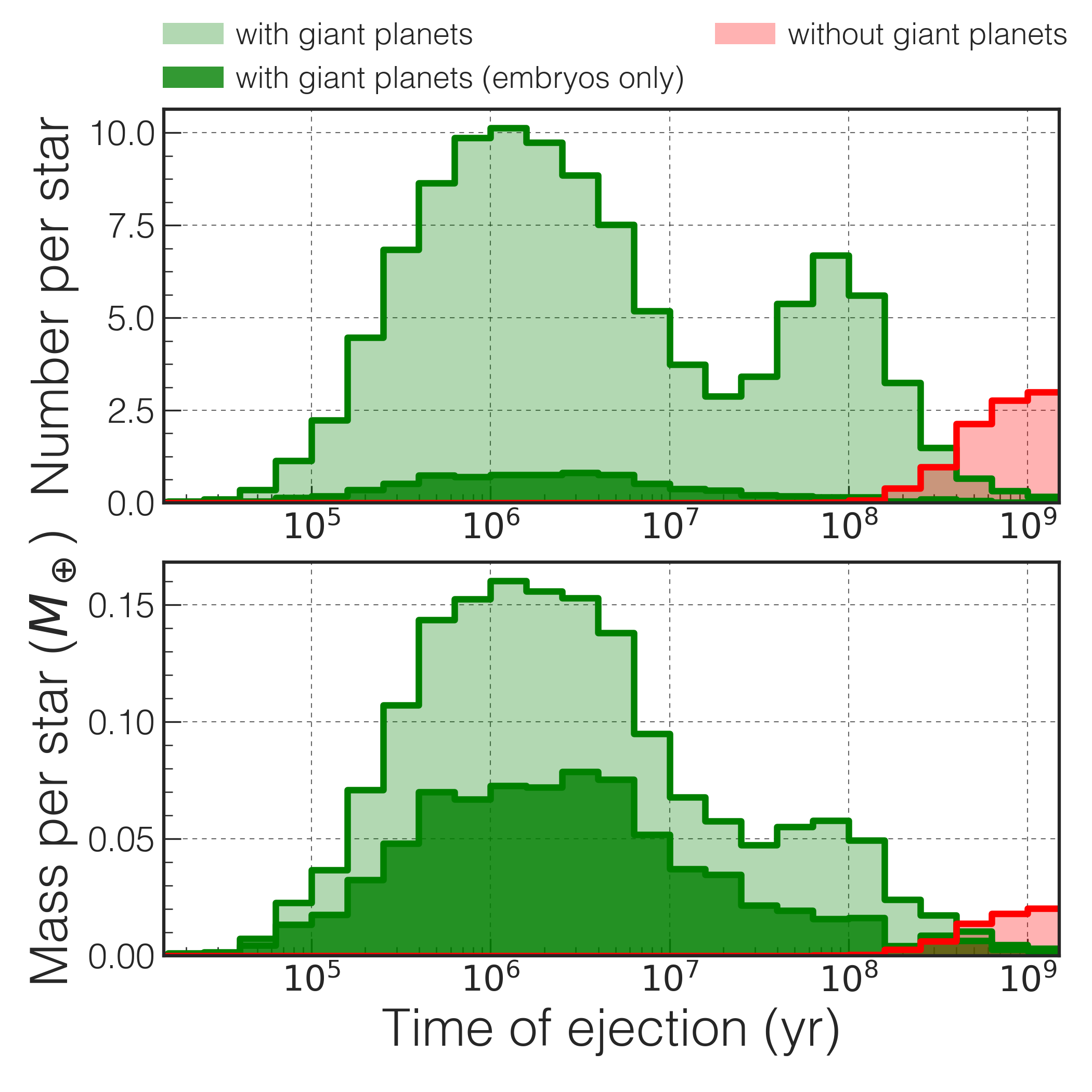}
\caption{Distributions of ejection times for simulations with giant planets (green) and without giant planets (red) shown as a function of the number of bodies ejected per star (upper panel) and the mass per star. The bin-widths are logarithmic. The dark green bins show the bodies we define as planets. The number of ejections in simulations with giant planets has two peaks in log-space, at about 10$^6$ Myr and 10$^8$ Myr. Ejections occur frequently early on in systems with giant planets, and the number of ejections decreases with time. Without giant planets, ejections are rare for the first several hundred million years, but thereafter the number of ejections occurs at a relatively constant rate, dropping by a factor of two from 0.5--2 Gyr. The bimodal distribution of ejections from the upper panel is evident in the lower panel, with ejections of both planets and planetesimals comprising the first peak around 10$^6$ Myr and only ejections of low-mass planetesimals and fragments from collisions in the second peak.
\label{fig:ejectiontime}}
\end{figure}

The time when ejections occur also differs radically between the simulations with and without giant planets. Figure~\ref{fig:ejectiontime} shows distributions of the ejection times for both sets in log-space. The simulations with giant planets (green histograms) can be approximated by a mixture of two log-normal distributions. The first log-normal from Figure~\ref{fig:ejectiontime} peaks at around 10$^6$ yr and a second peak appears around 10$^8$ yr. The red histograms in this plot show the simulations without giant planets. The first ejections don't occur until after 10$^6$ yr, peaking at around 500 Myr before a slow decline. With giant planets, very few ejections occur after several hundred million years, in contrast to the simulations without giant planets where ejections are still occurring at the end of the 2 Gyr simulations, albeit at a rate of $\sim$0.5 ejections of a planetesimal/fragment per 100 Myr. 

The first peak in the distribution of mass ejected as a function of ejection times (lower panel in Figure~\ref{fig:ejectiontime}) is still present and is the result of ejections of both planetesimals and planets. The second peak is much less pronounced in mass density compared with number density, with few planets being ejected at this time. Without giant planets, the low mass of the ejected bodies is evident.

\begin{figure}
\plotone{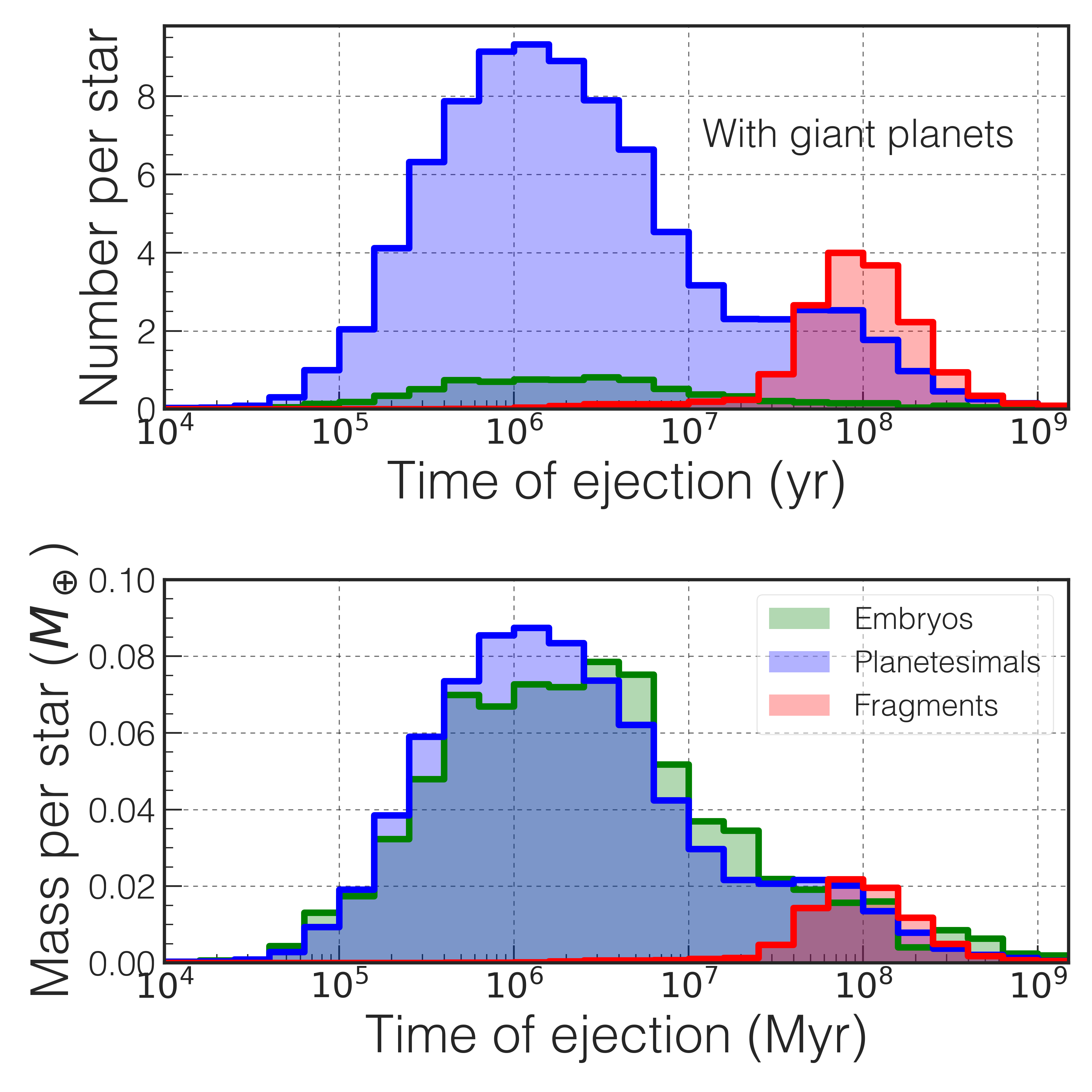}
\caption{For the simulations with giant planets, the types of bodies that are ejected are shown here. The upper panel shows the number of bodies ejected per logarithmic time bin, and the lower panel shows the mass ejected on the same timescale. Embryos (large bodies) are shown in green and are characterized by early ejections. Planetesimals are shown in blue and show an ejection timescale that matches the embryo ejection time but has a longer tail. Fragments are the product of collisions and are ejected at later times (red), presumably because they are reaccreted relatively quickly while the planets are still growing, whereas fragments that are created late face fewer encounters and larger perturbations. 
\label{fig:bodytype}}
\end{figure}

In Figure~\ref{fig:bodytype} we look at the types of bodies ejected as a function of time for the simulations with giant planets. The upper panel shows the number density and the lower panel shows the mass density. As we suspected from the lower panel of Figure~\ref{fig:ejectiontime}, we see that the second peak in number density is primarily due to a population of fragmented material. As fragments are created early on in the simulations, those created within the first 20 Myr are likely reaccreted by the embryos and planetesimals fairly quickly, whereas later, when the planets are mostly formed, any material that results in fragmentation can be more easily perturbed out of the system. The mass density of material is roughly evenly split between planets, planetesimals and fragments at the time of the second peak. The peak time of ejection for planetesimals is a little earlier than for planets, but the tails of the distributions are comparable.


While the number of Mars-sized free-floating planet detections predicted in this work is not inconsistent with previous predictions, we predict a severely depressed occurrence in Earth-sized planets. In the simulations with giant planets most bodies are ejected in the first tens of millions of years. This timescale explains the mass distribution of the material ejected: with most of the ejections occurring early in the simulations, there simply isn't the time required to build larger planets via pair-wise accretion. By the time Earth-like planets have largely finished forming, 30--100 Myr or so \citep{quintana16}, the dynamical interactions that cause planetary ejections are scarce. The two timescales of ejections shown in Figure~\ref{fig:bodytype} reflect different populations of ejected material. The first ejected population can be thought of as a primordial material - the planetesimals and embryos that were the remnants of star and giant planet formation. A second population of debris from collisions and original planetesimals that have experienced impacts themselves are ejected at this later time with on average less than one massive body being ejected from the system after about 200 Myr.

\begin{figure}
\includegraphics[width=0.47\textwidth]{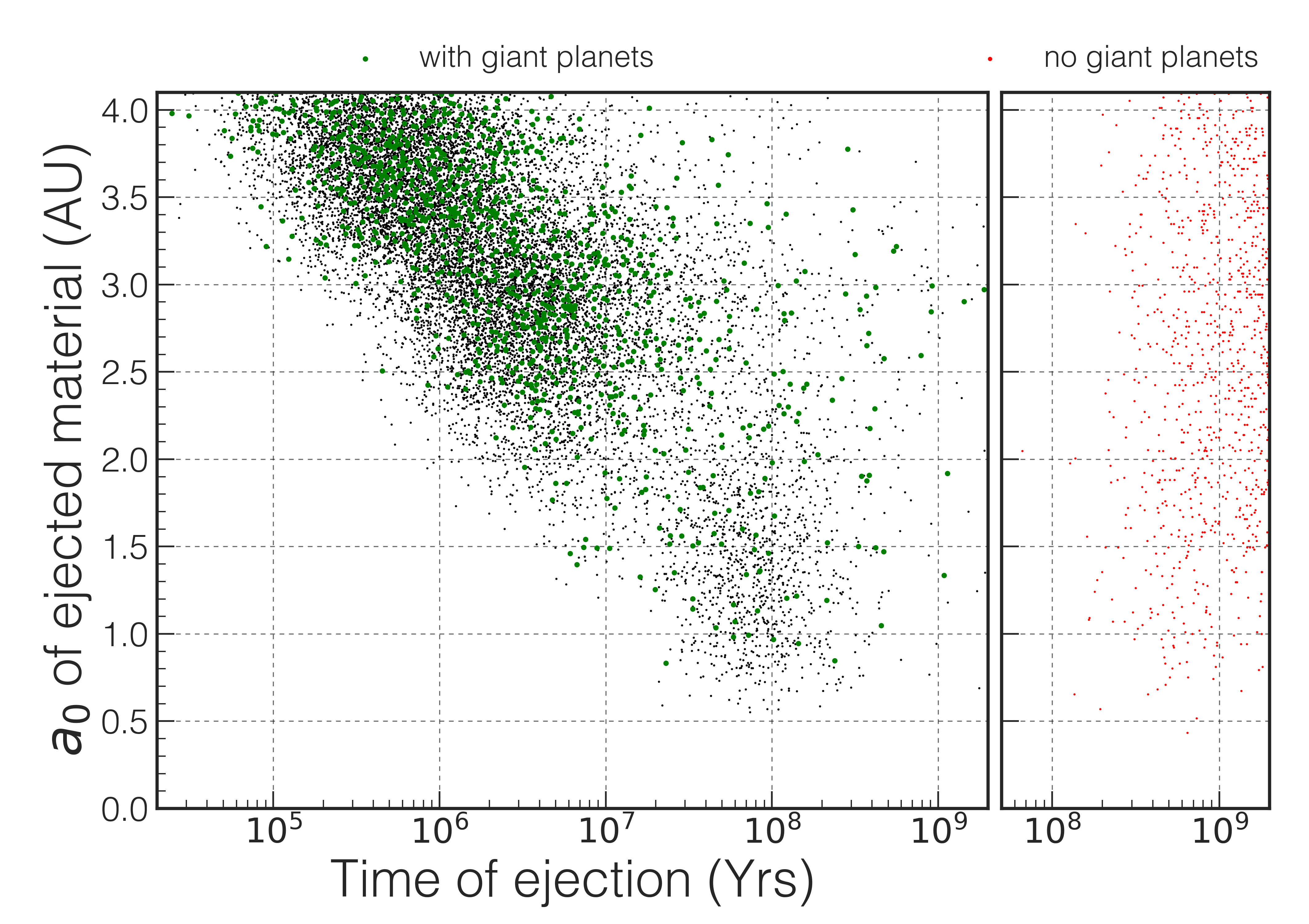}
\caption{The time when a body is ejected plotted against the initial semimajor axis of the body ($a_0$). In simulations with giant planets (left panel) planets are the larger green points and planetesimals are the small black dots. Bodies closest to Jupiter and Saturn are ejected first, beginning with ejected bodies at 4.0 AU approximately 100,000 years into the simulation. This distribution of ejected material follows a power law. There are also two distinct clusters with a population that began exterior to 2.0 AU being ejected first, followed by bodies that began inside 2.0 AU which are ejected later. This closer-in population is mostly reprocessed material that has undergone high-energy collisions. We have added a small jitter to the planet's $a_0$ for ease of visualization. When no giant planets are present (red dots, right panel), ejections don't start until around 100 Myr into the simulation and do not show a dependence on initial semi-major axis.
\label{fig:semitime}}
\end{figure}

The first bodies to be ejected from the simulations with giant planets are those that had initial orbital distances closest to the orbits of the giant planets. The blue dots in Figure~\ref{fig:semitime} show the initial position of all the ejected bodies and the time when they were ejected. This plot demonstrates that the time when ejections occur follows a power-law with the first ejections consisting of material close to 4.0 AU and occurring at around 100,000 yrs into the simulation. The ejection times also show two distinct clusters with a break at an initial semimajor axis of 2.0 AU. The outer material is ejected first and is primarily primordial material while the inner population is material that has been reprocessed through collisions. 

In simulations without giant planets, ejections typically don't start until about 200 Myr into the simulations, primarily because ejecting a planet requires a significant amount of angular momentum to be imparted to the ejected body. It is only once relatively massive terrestrial planets have begun to form that ejection of the low-mass material is possible. The ejections occur roughly uniformly with time throughout the simulations, with perhaps a drop by a factor of four between 500 Myr and 2 Gyr. The ejected bodies, as shown in the right panel of Figure~\ref{fig:semitime}, do not show any strong dependence on initial orbital distance. The planetesimal disk is well mixed by 100 Myr and the ejected material has no `memory' of where it began.

\section{Alternative giant planet configurations}\label{sec:othergiants}

We have thus far only considered two giant planet configurations: one with Jupiter and Saturn near their present orbits, and another with no giant planets. This binary choice is clearly not representative of the huge diversity of planetary system already discovered but it does provide two relatively extreme cases. As an intermediate example, \citet{Quintana.Lissauer:2014} looked at the effect of a 10 Earth-mass planet in Jupiter's orbit (near 5.2 AU) on terrestrial planet formation, assuming perfect accretion during collisions, using the same initial disk as that presented here. They found that the total mass of material ejected was between that of their simulations with no giant planets and with Jupiter and Saturn, as expected. While these types of systems with outer companions (super-Earths and ice giants) could conceivably significantly increase the number of ejected planets, they are unlikely to change mass distribution of the ejected material. We next examine the effects of various outer giant planets, some of which become unstable, on a disk of material that extends beyond Jupiter's orbit.

\subsection{Simulations with unstable gas giants}
We complement our analysis of ejected material with a separate set of simulations from \citet{raymond11, raymond12}. The most important difference between this set of simulations and those presented in Section~\ref{sec:results} is that these simulations included three gas giants whose orbits often became unstable. This allows us to connect our results with a large population of extra-solar planetary systems. The broad observed eccentricity distribution of giant exoplanets \citep[e.g.][]{butler06,udry07} can be reproduced if the vast majority (at least 75-90\%) of systems containing giant exoplanets go unstable \citep{moorhead05,juric08,chatterjee08,raymond10}. Instabilities are characterized by a series of scattering events between giant planets, and the surviving planets' stretched-out orbits are essentially scars left behind by this process \citep{rasio96,weidenschilling96}. 

The initial setup of the simulations from \citet{raymond11, raymond12} was divided into three zones. The inner zone -- corresponding to the terrestrial planet region -- was similar in structure to our other simulations, but included more total mass. The inner disk was populated by roughly 50 planetary embryos and 500 planetesimals. A total of 9 $M_\oplus$ was equally divided between the embryos (0.05-0.12 $M_\oplus$ each) and planetesimals (0.009 $M_\oplus$ each) and distributed following an $r^{-1}$ surface-density profile. 

In the middle zone, exterior to the terrestrial planet region, three giant planets were included in each simulation. The innermost gas giant was placed at 5.2 AU, and two additional giants were placed farther out. The spacing between adjacent gas giants was 4-5 mutual Hill radii, placing the planets in a marginally stable configuration \citep{chambers96,marzari02}. The gas giant masses varied among the different sets of simulations \citep[see][]{raymond12}. In the fiducial case, planet masses were chosen following their observed distribution \citep[$\frac{dN}{dM} \propto M^{-1.1}$; e.g.][]{butler06} between 1 Saturn mass and 3 Jupiter masses. 

A 10 AU wide belt containing 50 $M_\oplus$ in 1000 planetesimals was placed exterior to the gas giants. The inner edge of the belt was 4 Hill radii beyond the outermost gas giant. This setup was inspired by models for the Solar System's primordial planetesimal disk, which may have been disrupted by a late instability in the giant planets' orbits \citep[i.e., the ``Nice model'' of][]{tsiganis05,morbidelli07}. 

The total batch from \citet{raymond11, raymond12} contains roughly 500 simulations. For simplicity, we focus on the fiducial set of simulations, which has the advantage of being well characterized and reproducing the giant exoplanets' eccentricity distribution with no free parameters. 
In 96 out of 152 simulations (63\%) the giant planets went unstable but in the other 56 simulations they remained stable. The survival or destruction of terrestrial planets in unstable systems was linked directly to the strength of the giant planet instability, which can be measured in terms of either the closest approach of a giant planet during the scattering phase or by the eccentricity of the surviving giant planets \citep{raymond11, raymond12}, in particular of the innermost giant.

\begin{figure}
\includegraphics[width=0.47\textwidth]{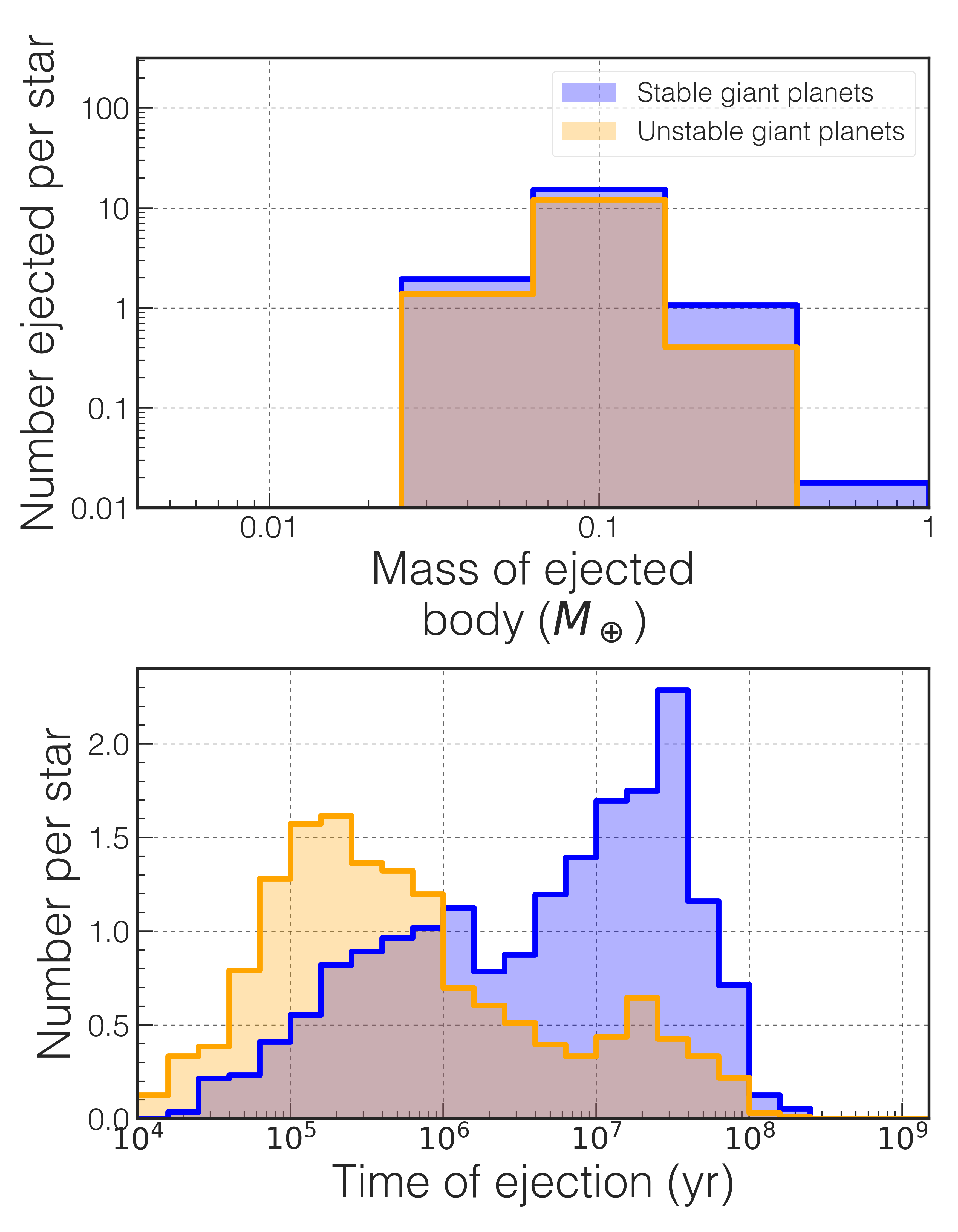}
\caption{The mass of embryos ejected from the 152 simulations of \citet{raymond11} (upper panel) and the time of ejection (lower panel). The body masses at the start of the simulations were between 0.05 and 0.12 $M_\oplus$, and most ejected bodies did not grow before being ejected. The two colors are linked with the giant planets' evolution in the system in which that particular body was ejected. Blue represent systems in which the giant planets' orbits remained stable and are comparable to the simulations with giant planets presented in this paper. Unstable systems are shown in orange and provide a different formation outcome. That mass distribution in these simulations is broadly consistent with the other simulations presented here.
\label{fig:mejec_tejec}}
\end{figure}

Figure~\ref{fig:mejec_tejec} shows the mass distribution and the time when bodies were ejected from the 152 fiducial simulations. The clump in mass between 0.05 and 0.12 $M_\oplus$ corresponds to the initial distribution of embryo masses. The relatively small number of ejected bodies with a mass above this indicates that most ejected embryos were primordial, meaning that they had not undergone any large collisions between the start of the simulation and the time of their ejection. 

Although systems with unstable gas giants are far more destructive in general, they ejected somewhat fewer embryos than systems with stable gas giants. In systems with unstable gas giants, the median number of ejected embryos was 14, versus a median of 18 embryos ejected in the simulations with stable gas giants. The unstable systems ejected a mean of 1.4 $M_\oplus$ in embryos per system vs. 1.9 $M_\oplus$ per stable system. In contrast, a median of 23 embryos collided with the central star in the unstable systems, versus a median of just 1 embryo colliding with the star in stable systems.

The bodies ejected from unstable systems can be divided by the time of ejection relative to the start of the giant planet instability. Given that there was a distribution of instability timescales, this does not translate into an obvious absolute timescale. In these simulations, the median instability timescale was roughly $10^5$ years, with a tail extending out to $\sim$100 Myr. As shown in Figure~\ref{fig:mejec_tejec}, ejections are common prior to the giant planets becoming unstable; the instability itself is not a significant driver of ejection, although by the nature of having a dynamically excited system seems to skew ejections to earlier times. Nonetheless, the most massive ejected planet in Fig.~\ref{fig:mejec_tejec} was a 1.04 $M_\oplus$ planet ejected during an instability. 

The results from these simulations -- in particular those with stable giant planets -- are broadly consistent with those from the simulations from Section~\ref{sec:results} with giant planets on Solar System-like orbits. Specifically, the embryos that were ejected tended to be between one and two Mars-masses. The vast majority of ejected embryos had undergone no accretion, and only a small fraction had grown to more than twice their initial mass. The most massive embryo ejected in a simulation with stable giant planets was 0.54 $M_\oplus$. Despite the differences in numerical resolution, embryo mass, and integration method, the concordance between this set of simulations and those presented in Section~\ref{sec:results} inspires confidence in our results.

\section{Estimating the Galactic terrestrial-mass free-floating population}\label{sec:galactic}

The contrast between properties of the ejected material in the three different simulations is intriguing. If giant planets are ubiquitous, then there exists a very significant population of Mars-sized material to make up a population of free-floating planets -- here we show as many as 8 per star with giant planets in the Galaxy from inner planetary systems (c.f. Table~\ref{tab:results}) and about double the number from larger orbital distances. However, there are strong indications that the occurrence rate of giant planets is significantly lower than unity; \citet{wittenmyer16} estimate an occurrence of just 6.2\% of giant planets between 3 and 7 AU from their stars. This low occurrence doesn't seem particularly outlandish, particularly considering that the most common spectral type -- M-dwarfs -- very rarely host giant planets \citep{dressing15,clanton14,clanton16}. 

To apply our results to the Galactic population, we divide the known systems of exoplanets into three categories (in order of increasing abundance). 


\begin{enumerate}
  \item Systems like our own, with gas giants on wide, near-circular orbits. Estimates of the occurrence rates for giant planets range from 5--20\% with perhaps a third of these giant planets on wide, low-eccentricity orbits \citep[e.g.][]{butler06,zakamska11}. We assume that this subset makes up 6\% of stars in the Galaxy.
  \item Systems with gas giants on eccentric orbits. The remainder of the giant exoplanets have orbits that are either too close-in or too eccentric to be considered Jupiter-like. This comprises around two-thirds of the entire giant exoplanet population, approximately 12\% of stars.
  \item Systems without gas giants. Most stars have no gas giants. Of course, a large fraction may host super-Earths or ice giants \citep{mayor11,howard10,howard12,fressin13}, but, as we showed in Section~\ref{sec:implications}, we do not expect these systems to contribute significantly to the population of rocky free-floating objects. 
\end{enumerate}

For each class of systems, one of our sets of simulations allows us to crudely estimate the rate of ejection of rocky planet-sized bodies (with the various caveats discussed in Section~\ref{sec:limitations}). We can then produce a simple linear combination of these estimates, weighted by the occurrence rate of each type of system, e.g.
\begin{align}
\eta = &F_\textrm{giants, circular} \times FF_\textrm{giants, circular}\nonumber\\
&+ F_\textrm{giants, eccentric} \times FF_\textrm{giants, eccentric}\nonumber\\
&+ F_\textrm{no giants} \times FF_\textrm{no giants}
\end{align}
where $\eta$ is the number of terrestrial-mass free-floating planets ejected per star, $F$ is the fraction of stars falling into the three categories (circular giants, eccentric giants, and no giant planets), and $FF$ is the number of free-floating planets ejected from that category. We do not include systems with hot Jupiters in this estimate because they only account for approximately 1\% \citep{Wright12} of stars and therefore are unlikely to significantly affect the overall occurrence rate. 

Our results from Section~\ref{sec:results} demonstrated that systems without gas giants are unlikely to contribute to the population of free-floating rocky planets. In our simulations of the inner Solar System we found that approximately 8 Mars-mass planets are ejected per simulation, while in the simulations from \citet{raymond11,raymond12} about twice this number of Mars-mass bodies are ejected (likely because they use a more massive initial disk), so we will take the average of these simulations. Finally, in the simulations with unstable gas giants, about an average of 14 Mars-mass embryos were ejected. So, this would imply $(0.5 * (8+18))\times0.06$ bodies in stable simulations, $14\times0.12$ in unstable simulations for a total of 2.5 Mars-mass embryos per star in the Galaxy. Owing to relatively poor constraints on in the mass in the solids in the terrestrial region of the initial disk, this estimate probably has an uncertainty of a factor of several.

\section{Implications for WFIRST}\label{sec:implications} 
\begin{figure}
\plotone{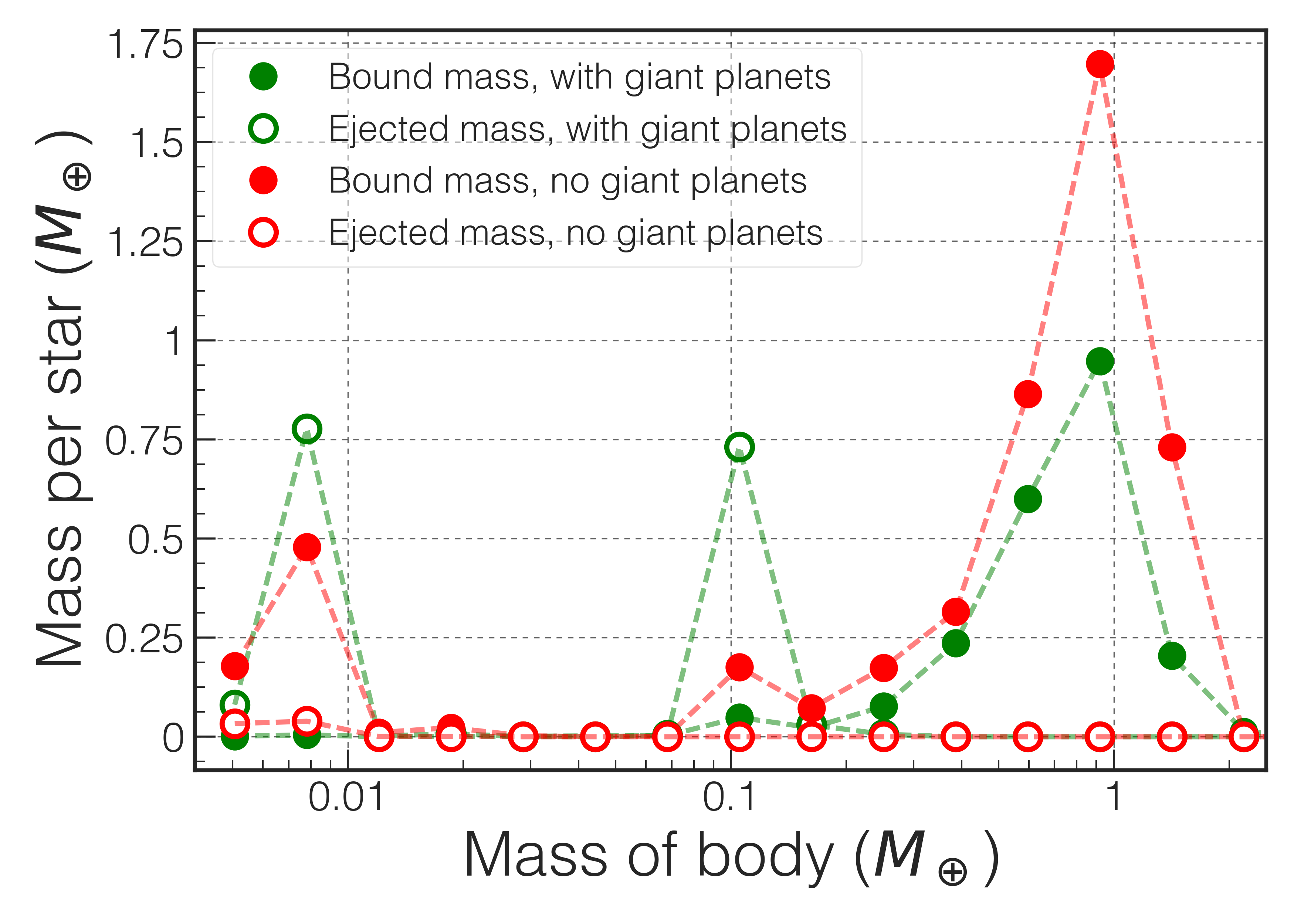}
\caption{A comparison between the mass distribution of bound planets in our simulations with the planets that were ejected. While both types of simulations form bound planets with masses from 0.3--1.0 $M_\oplus$, no large bodies are ejected. The properties of the bound population do not match the ejected population. Specifically, more planets form without giant planets and more planetesimals remain bound. We only show data from the primary simulations in this figure and do not include data from \citet{raymond11}.
\label{fig:bound}}
\end{figure}

Previous estimates of the number of detections from WFIRST have primarily been based on the population of bound planets, either the Solar System population or the observed population from microlensing \citep{cassan12}. Figure~\ref{fig:bound} shows a comparison between the mass distribution of bound and ejected planets for simulations with (green) and without (red) giant planets from our simulations of the inner planetary systems. Both simulation types show peaks in the bound planet mass distribution at around one Earth-mass, while there is no ejected material more massive than 0.3 $M_\oplus$. A clear takeaway from this work is that the bound population is radically different from the ejected population. 

We estimated 2.5 free-floating terrestrial-mass planets per star in the Galaxy (c.f. Section~\ref{sec:galactic}). Now we look at how many of these planets would be detectable by the microlensing component of the WFIRST mission. The sensitivity of WFIRST to detecting microlensing events was estimated by \citet{spergel15}, who show estimates of the free-floating population based on simulations performed by \citet{raymond11}. 
\citet{spergel15} predict that with a single ejected planet per star, WFIRST would find 5.7 Mars-sized free-floating planets. Taking this estimate from WFIRST and combining it with our simulation results, we find a number three times greater at 14.3 planets detected by WFIRST. While this estimate is based on relatively simplistic assumptions, it is suggestive of a significant number of microlensing events in WFIRST data.

\section{Limitations and the effect of initial conditions}\label{sec:limitations}

An important consideration for all studies of this type is the effect of the initial conditions on the final results. Parameters such as the total disk mass, size, and mass distribution of protoplanets and the presence and proximity of outer giant planet companions can weigh heavily on the final results. Because these types of $N$-body simulations are computationally intensive and therefore a large number of simulations are needed due to their stochastic nature, a full exploration of parameter space for all initial conditions is not feasible. We therefore limit our primary study of ejected material from the terrestrial region (Section~\ref{sec:results}) to a single star/disk model that is motivated by decades of research on Solar System formation \citep{Wetherill:1994,Chambers:2001,Raymond.etal:2004,Obrien.etal:2006,Raymond.etal:2009,Quintana.etal:2002,Quintana.Lissauer:2014} and vary the presence of giant planets.



There is very likely a great diversity of disk mass distributions, however, that result from the star formation process, as evidenced by studies of Kepler planetary systems \citep{moriarty15,RaymondCossou2014,quintana14}. While ALMA observations are revolutionizing our understanding of the earlier stages of protoplanetary disks \citep[e.g.][]{hltau}, we have relatively poor constraints on the total mass and distribution of solids in disks that are old enough to have lost their gas component. However, there is evidence that our radial distribution is at least plausible \citep{gilloteau11,mathews13}. Simulations using the minimum mass Solar nebula model, which our disk is based on, have been successful at reproducing systems of terrestrial planets with properties comparable to those in our Solar System. Many numerical studies have explored the effects of varying the disk with these properties on the final planets that form. While these studies have used perfect-accretion $N$-body models, they are still valuable for getting a sense of the parameters that have stronger effects on the abundance and timescale of ejected mass. \citet{Raymond.etal:2005b} examined the impact of varying the disk surface-density profile on the total mass ejected in systems with Jupiter and Saturn and found a difference of about a factor of three in ejected mass between very shallow and very steep (compared to the minimum mass Solar nebula model) profile disks. We use an intermediate slope surface-density profile which is unlikely to bias the total ejected mass by more than a factor of about two and is therefore unlikely to be the dominant source of uncertainty in our estimates.

\citet{kokubo06} showed that for a given total disk mass, the properties of the final planets that form are not highly sensitive to the number, mass, or bulk density of the planetary embryos that compose the disk. Variations in total disk mass, however, do influence the types of planets that form. More massive disks tend to form fewer, more massive planets \citep{kokubo06,raymond07b,Hansen:2013}, although the timescale problem for forming and ejecting Earth-mass planets still applies. With more massive disks having more material available to eject, our predictions on the number of free-floating planets that WFIRST will detect is fairly conservative. If significantly more massive disks than we have used are the norm, we would expect to see more Mars-sized planets.



The bimodal mass distribution that we consider in our initial disk is the result of a runaway and oligarchic growth phase of protoplanets during the earlier stages of planet formation \citep{Kokubo.Ida:1998}. Prior to runaway growth, the sizes of protoplanets in the disk are comparable and their eccentricities and inclinations remain low due to dynamical friction. While the abundance of ejected material during these earlier stages is difficult to estimate, the bodies ejected during this epoch are likely too small to be detectable (they only reach Mars-size in the post-oligarchic growth phase).

The sizes of our initial planetesimals ($\sim$1 lunar-mass, or 0.009 $M_{\oplus}$) and the minimum fragment size ($\sim$0.4 lunar-mass, or 0.005 $M_{\oplus}$) were chosen to keep our simulations computationally tractable while retaining enough bodies to mimic a size distribution that provides dynamical friction to the embryos and final planets. Therefore, we do not draw any conclusions on the specific size distribution of ejected bodies smaller than the Moon, we class all this mass as `small' and treat this set as a whole. \citet{Obrien.etal:2006} used the same initial disk mass and surface-density profile as our model, but with a factor of four more planetesimals. They found that a near identical fraction of the initial mass was ejected from their simulations as those presented here. We therefore conclude that the resolution of the planetesimal population does not strongly affect the abundance of the ejected mass. 

Our simulations run for 2 Gyr. While in the giant planet simulations formation and evolution of the terrestrial planets is essentially complete, the simulations with no giant planets show ongoing activity. We opted not to continue these simulations beyond 2 Gyr because just running this far took several months. However, we find it unlikely that the evolution of these system will vary dramatically over the next 2--10 Gyr. There may be more collisions, but it is unlikely to cause a significant change in the mass distribution of the ejected material or the ejection frequency.


Finally, we do not consider external forces such as passing stars that can dynamically drive ejections. However, although passing stars in dense clusters can stir up material and cause planets to become unstable, most planetary systems do not experience stars passing close enough to have a significant impact on the inner region of planetary systems where terrestrial planets form \citep{laughlin00,li15}.

\section{Conclusions}\label{sec:conclusions}

In this study we focus on quantifying the mass that is ejected from a star/disk system during the planet formation process using state-of-the-art numerical $N$-body simulations. Our primary study involves running hundreds of simulations for two sets of initial configurations: a disk of protoplanets around a Sun-like star with and without outer giant planets analogous to Jupiter and Saturn. We use the same initial disk of protoplanets for all simulations, but with small changes in the initial conditions of one body to account for chaos. Our disk model is based on the distribution of solids that are thought to be present about 10 Myr after the birth of our Solar System, an epoch in which the gas component of the disk has dispersed, giant planets are formed, and Moon-to-Mars-sized protoplanets have accreted \citep[][and references therein]{quintana16}. Specifically, our disk includes about 5 $M_\oplus$ of solids in 26 Mars-sized ($\sim$0.1 $M_\oplus$) embryos and 260 Moon-sized planetesimals ($\sim$0.01 $M_\oplus$) all within 4 AU from a Sun-like star. The goal of these simulations is to estimate the abundance and mass distribution of free-floating planets that originated in the terrestrial region (i.e., `Earth-like'). Our simulations were performed using an $N$-body integrator that allows both accretion and fragmentation \citep{Chambers:2013} and we set the resolution (minimum allowed mass during fragmentation) to 0.4 lunar-mass (0.005 $M_\oplus$).

We find that in simulations with giant planets, roughly half the ejected material is composed of bodies with masses greater than 0.06 $M_\oplus$ and half in smaller bodies. While many Mars-mass bodies were ejected during the final stages of planet formation that we simulate, no planets more massive than 0.3 $M_\oplus$ were ejected. The primary reason is that nearly all ejections occur within the first 30 Myr, which is prior to the giant-impact phase of planet formation when planets accrete enough material to grow to Earth size. Most of the ejections essentially mirrored our initial bimodal disk mass distribution, but with embryos that only had time to grow to about one-third of an Earth mass and many planetesimals and fragments that didn't accrete a substantial amount of mass.

With no giant planets, all the ejected mass is in planetesimals and amounts to just 1\% of the initial disk. The vast majority of solids in the disk remains bound to the star throughout our 2 Gyr simulations. The material that is ejected does not leave the system until Earth-mass planets have had time to form, contrary to the set of simulations with giant planets. However, without giant planet perturbations, the systems lack a mechanism to impart enough angular momentum to cause planet ejections. This implies that stars that lack outer giant planet companions contribute very little to the free-floating planet population.


We also examined ejected mass from simulations performed previously \citep{raymond11, raymond12} that explored a different parameter space than our initial study. About 150 simulations were presented that included a disk with about 9 $M_\oplus$ of solids in the terrestrial region (so almost double the mass of our primary study), three giant planets (one in Jupiter's orbit, the others exterior), and a population of 50 $M_\oplus$ of solids in 1000 planetesimals exterior to the giant planets and out to 10 AU. This set allowed us to examine the effects of different giant planet architectures, giant planet instabilities, and a more massive disk on the population of ejected material. While these simulations include more mass in embryos, a wider disk, and additional giant planets, their results on the mass distribution of ejected embryos are consistent with the population ejected from the terrestrial region in our primary study. Interestingly, the abundance of embryos ejected in simulations where the giant planets became unstable was lower than when the giant planets remained stable, primarily because the instabilities tended to perturb a bulk of the lost material into the star rather than into interstellar space.
With the exception of two objects, all of the ejected bodies in this set of simulations had masses lower than 0.4 $M_\oplus$.

Taken together, these simulation imply that few Earth-mass free-floating planets are ejected during the planet formation process. The population of free-floating planets is likely composed of Mars-sized bodies, many of which may have originated from the outer cooler regions of a protoplanetary disk.

Our results suggest that the abundance of terrestrial planets that originate from the terrestrial region ($<$4 AU) detected by WFIRST will depend on the frequency of giant planets. We combined the results from all simulations with the occurrence rates of different giant planet architectures and stellar types to estimate a Galactic population. We predict that roughly 2.5 terrestrial-mass planets are ejected per star. However, the vast majority of these come from the $<20$\% of stars that host giant planets.

The WFIRST microlensing program will be sensitive to planets less massive than Mars. We predict that WFIRST will find approximately 15 Mars-mass free-floating planets, and the number could be even larger if typical disk masses are significantly more massive than we used in this work.






\acknowledgments{
E.V.Q.'s research was partially supported by a NASA Postdoctoral Program Senior Fellowship at the NASA Ames Research Center, administered by Universities Space Research Association under contract with NASA. The simulations presented here were performed using the Pleiades Supercomputer provided by the NASA High-End Computing (HEC) Program through the NASA Advanced Supercomputing (NAS) Division at Ames Research Center. The authors would like to thank Chris Henze for his expertise and assistance in performing simulations on the Pleiades Supercomputer.
}

\bibliographystyle{apj}
\bibliography{refstom2}

\end{document}